\documentclass[preprint,12pt]{elsarticle}

\usepackage{graphicx}

\usepackage{amsmath}
\usepackage{amssymb}
\usepackage{amsthm}

\journal{J.Comp.Phys.}

\begin{document}

\begin{frontmatter}



\title{A Spectral Analysis Method for Automated Generation of Quantum-Accurate Interatomic Potentials}


\author[snl1]{A.P.~Thompson}\corref{cor1}
\ead{athomps@sandia.gov}
\author[snl2]{L.P.~Swiler}
\ead{lpswile@sandia.gov}
\author[snl3]{C.R.~Trott}
\ead{crtrott@sandia.gov}
\author[snl4]{S.M.~Foiles}
\ead{foiles@sandia.gov}
\author[snl4,drexel]{G.J. Tucker}
\ead{gtucker@coe.drexel.edu}

\cortext[cor1]{Corresponding author}
 
\address[snl1]{Multiscale Science Department, Sandia National Laboratories, P.O. Box 5800, MS 1322, Albuquerque, NM 87185}
\address[snl2]{Optimization and Uncertainty Quantification Department, Sandia National Laboratories, P.O. Box 5800, MS 1318, Albuquerque, NM 87185}
\address[snl3]{Scalable Algorithms Department, Sandia National Laboratories, P.O. Box 5800, MS 1322, Albuquerque, NM 87185}
\address[snl4]{Computational Materials and Data Science Department, Sandia National Laboratories, P.O. Box 5800, MS 1411, Albuquerque, NM 87185}
\address[drexel]{Materials Science and Engineering Department, Drexel University, Philadelphia, PA 19104}

\begin{abstract}
We present a new interatomic potential for solids and liquids called Spectral Neighbor Analysis Potential (SNAP).
The SNAP potential has a very general form and uses machine-learning techniques to reproduce the 
energies, forces, and stress tensors of a large set of small configurations of atoms, which are obtained 
using high-accuracy quantum electronic structure (QM) calculations.
The local environment of each atom is characterized by a set of bispectrum components of the local
neighbor density projected on to a basis of hyperspherical harmonics in four dimensions. 
The bispectrum components are the same bond-orientational order parameters  
employed by the GAP potential~\cite{Bartok2010}.  The SNAP potential, unlike GAP, assumes
a linear relationship between atom energy and bispectrum components.  
The linear SNAP coefficients are determined using weighted least-squares linear regression against the
full QM training set.   This allows the SNAP potential to be fit in a robust, automated manner to large QM data 
sets using many bispectrum coefficients. The calculation of the bispectrum components and the SNAP potential
are implemented in the LAMMPS parallel molecular dynamics code. We demonstrate that a previously
unnoticed symmetry property can be exploited to reduce the computational cost of the force calculations
by more than one order of magnitude.   
We present results for a SNAP potential for tantalum, showing that it accurately reproduces a range of
commonly calculated properties of both the crystalline solid and the liquid phases.  In addition, unlike simpler 
existing potentials, SNAP correctly predicts the energy barrier for screw dislocation migration in BCC tantalum. 

\end{abstract}


\end{frontmatter}



\section{Introduction}
\label{sec:intro}

Classical molecular dynamics simulation (MD) is a powerful approach for describing the mechanical, chemical, 
and thermodynamic behavior of solid and fluid materials in a rigorous manner~\cite{Griebel2007}. 
The material is modeled as a large collection of point masses (atoms) whose motion is tracked by integrating 
the classical equations of motion to obtain the positions and velocities of the atoms at a large number of timesteps.  
The forces on the atoms are specified by an interatomic potential that defines the potential energy of the 
system as a function of the atom positions. Typical interatomic potentials are computationally inexpensive 
and capture the basic physics of electron-mediated atomic interactions of important classes of materials, 
such as molecular liquids and crystalline metals. Efficient MD codes running on commodity workstations 
are commonly used to simulate systems with $N = 10^5 - 10^6$ atoms, the scale at which many interesting 
physical and chemical phenomena emerge.   
Quantum molecular dynamics (QMD) is a much more computationally intensive method for solving a similar
physics problem~\cite{Schultz2005}.  
Instead of assuming a fixed interatomic potential, the forces on atoms are obtained by explicitly solving 
the quantum electronic structure of the valence electrons at each timestep. Because MD potentials are short-ranged, 
the computational complexity of MD generally scales as $O(N)$, whereas QMD calculations require global self-consistent 
convergence of the electronic structure, whose computational cost is $O(N^\alpha_e)$, where $2 < \alpha < 3$ 
and $N_e$ is the number of electrons.  For the same reasons, MD is amenable to spatial decomposition on 
parallel computers, while QMD calculations allow only limited parallelism.  

As a result, while high accuracy QMD simulations have supplanted MD in the range $N =10-100$ atoms, 
QMD is still intractable for $N > 1000$, even using the largest  supercomputers.  Conversely, typical 
MD potentials often exhibit behavior that is inconsistent with QMD simulations. This has led to great 
interest in the development of MD potentials that match the QMD results for small systems, but can still be 
scaled to the interesting regime $N = 10^5 - 10^6$ atoms~\cite{Bartok2010, Artrith2012, Li2003}.  
These quantum-accurate potentials require many more 
floating point operations per atom compared to conventional potentials, but they are still short-ranged. 
So the computational cost remains $O(N)$, but with a larger algorithm pre-factor. 

In this paper, we present a new quantum-accurate potential called SNAP.  It is designed to model the 
migration of screw dislocations in tantalum metal under shear loading, the fundamental process underlying plastic deformation 
in body-centered cubic metals. In the following section we explain the mathematical
structure of the potential and the way in which we fit the potential parameters to a database of quantum electronic
structure calculations.  We follow that with a brief description of the implementation of the 
SNAP potential in the LAMMPS code.  We demonstrate that a previously
unnoticed symmetry property can be exploited to reduce the computational  cost of the force calculations
by more than one order of magnitude.   
We then present results for the SNAP potential that we have developed
for tantalum.  We find that this new potential accurately reproduces a range of properties of solid and liquid
tantalum.  Unlike simpler potentials, it 
correctly matches quantum MD results for the screw dislocation core structure and minimum energy pathway for displacement 
of this structure, properties that were not included in the training database.

\section{Mathematical Formulation}
\label{sec:formulation}

\subsection{Bispectrum coefficients}
\label{subsec:bispectrum}

The quantum mechanical principle of near-sightedness tells us that the electron density at a point is only weakly affected by atoms that are not near.  This provides support for the common assumption that the energy of a configuration of atoms is dominated by contributions from clusters of atoms that are near to each other.  It is reasonable then to seek out descriptors of local structure and build energy models based on these descriptors.  Typically, this is done by identifying geometrical structures, such as pair distances and bond angles, or chemical structures, such as bonds.  Interatomic potentials based on these approaches often produce useful qualitative models for different types of materials, but it can be difficult or impossible to adjust these potentials to accurately reproduce known properties of specific materials.  Recently, Bart{\'{o}}k et al. have studied several infinite classes of descriptor that are related to the density of neighbors in the spherically symmetric space centered on one atom~\cite{Bartok2010, BartokThesis2010, Bartok2013}.  They demonstrated that by adding descriptors of successively higher order, it was possible to systematically reduce the mismatch between the potential and the target data.  One of these descriptors, the bispectrum of the neighbor density mapped on to the 3-sphere, forms the basis for their Gaussian Approximation Potential (GAP)~\cite{Bartok2010}.  We also use the bispectrum as the basis for our SNAP potential.  We derive this bispectrum below, closely following the notation of Ref.\cite{Bartok2013}. 

The density of neighbor atoms around a central atom $i$ at location $\textbf{r}$ can be considered as a sum of $\delta$-functions located in a three-dimensional space:

\begin{equation}
\rho_i ({\bf r}) = \delta({\bf r}) + \sum_{r_{ii'} < R_{cut}}{f_c(r_{ii'}) w_{i'} \delta({\bf r}-{\bf r}_{ii'})}
\end{equation}

where ${\bf r}_{ii'}$ is the vector joining the position of the central atom $i$ to neighbor atom $i'$.  The $w_{i'}$ coefficients are dimensionless weights that are chosen to distinguish atoms of different types, while the central atom is arbitrarily assigned a unit weight.  The sum is over all atoms $i'$ within some cutoff distance $R_{cut}$.  The switching function $f_c(r)$ ensures that the contribution of each neighbor atom goes smoothly to zero at $R_{cut}$.  The angular part of this density function can be expanded in the familiar basis of spherical harmonic functions $Y^l_m(\theta,\phi)$, defined for $l = 0,1,2,\ldots$ and $m = -l,-l+1,\ldots,l-1,l$~\cite{Varshalovich1987}.  The radial component is often expanded in a separate set of radial basis functions that multiply the spherical harmonics.  Bart{\'{o}}k et al. made a different choice, mapping the radial distance $r$ on to a third polar angle $\theta_0$ defined by,

\begin{equation}
\theta_0 = \theta_0^{max}\frac{r}{R_{cut}}
\end{equation}

The additional angle $\theta_0$ allows the set of points $(\theta, \phi, r)$ in the 3D ball of possible neighbor positions to be mapped on to the set of points $(\theta, \phi, \theta_0)$ that are a subset of the 3-sphere.  Points south of the latitude $\theta_0^{max}$ are excluded.  It is advantageous to use most of the 3-sphere, while still excluding the region near the south pole where the configurational space becomes highly compressed.

The natural basis for functions on the 3-sphere is formed by the 4D hyperspherical harmonics $U^j_{m,m'}(\theta_0,\theta,\phi)$, defined for $j=0,\frac{1}{2},1,\ldots$ and 
$m,m' = -j,-j+1,\ldots,j-1,j$~\cite{Varshalovich1987}.  These functions also happen to be the elements of the unitary transformation matrices for spherical harmonics under rotation by angle $2\theta_0$ about the axis defined by $(\theta, \phi)$.   When the rotation is parameterized in terms of the three Euler angles, these functions are better known as $D^j_{m,m'}(\alpha, \beta, \gamma)$, the Wigner $D$-functions, which form the representations of the $SO(3)$ rotational group~\cite{Meremianin2006, Varshalovich1987}.  Dropping the atom index $i$, the neighbor density function can be expanded in the $U^j_{m,m'}$ functions

\begin{equation}
\rho({\bf r}) = \sum_{j=0,\frac{1}{2},\ldots}^{\infty}\sum_{m=-j}^{j}\sum_{m'=-j}^{j} u^j_{m,m'} U^j_{m,m'}(\theta_0,\theta,\phi)
\end{equation}

where the expansion coefficients are given by the inner product of the neighbor density with the basis function. Because the neighbor density is a weighted sum of $\delta$-functions, each expansion coefficient can be written as a sum over discrete values of the corresponding basis function, 

\begin{equation}
u^j_{m,m'} = U^j_{m,m'}(0,0,0) + \sum_{r_{ii'} < R_{cut}}{f_c(r_{ii'}) w_{i'} U^j_{m,m'}(\theta_0,\theta,\phi)} 
\end{equation}

The expansion coefficients $u^j_{m,m'}$ are complex-valued and they are not directly useful as descriptors, because they are not invariant under rotation of the polar coordinate frame.  However, the following scalar triple products of expansion coefficients can be shown to be real-valued and invariant under rotation~\cite{Bartok2013}.

\newcommand{\hcoeff}[9]{H\!\!{\tiny\begin{array}{l}#1 #2 #3 \\ #4 #5 #6 \\ #7 #8 #9 \end{array}}}
  
\begin{equation}
B_{j_1,j_2,j}  = \\
\sum_{m_1,m'_1=-j_1}^{j_1}\sum_{m_2,m'_2=-j_2}^{j_2}\sum_{m,m'=-j}^{j} (u^j_{m,m'})^*
\hcoeff{j}{m}{m'}{j_1}{\!m_1}{\!m'_1}{j_2}{m_2}{m'_2}
u^{j_1}_{m_1,m'_1} u^{j_2}_{m_2,m'_2}
\end{equation}

The constants 
$\hcoeff{j}{m}{m'}{j_1}{\!m_1}{\!m'_1}{j_2}{m_2}{m'_2}$
are coupling coefficients, analogous to the Clebsch-Gordan coefficients for rotations on the 2-sphere. These invariants are the components of the bispectrum.  They characterize the strength of density correlations at three points on the 3-sphere.  The lowest-order components describe the coarsest features of the density function, while higher-order components reflect finer detail.   An analogous bispectrum can be defined on the 2-sphere in terms of the spherical harmonics.  In this case, the components of the bispectrum are a superset of the second and third order bond-orientational order parameters developed by Steinhardt et al.~\cite{Steinhardt1983}.  These in turn are specific instances of the order parameters introduced in Landau's theory of phase 
transitions~\cite{Landau1980}.

The coupling coefficients are non-zero only for non-negative integer and half-integer values of $j_1, j_2,$ and $j$ satisfying the conditions $\| j_1-j_2 \| \leq j \leq j_1 + j_2$ and $j_1+j_2 - j$ not half-integer~\cite{Meremianin2006}.  In addition, $B_{j_1,j_2,j}$ is symmetric in $j_1$ and $j_2$.  Hence the number of distinct non-zero bispectrum coefficients with indices $j_1, j_2, j$ not exceeding a positive integer $J$ is $(J+1)^3$. Furthermore, it is proven in the appendix that bispectrum components with reordered indices are related by the following identity:

\begin{equation}
\label{eqn:symm}
\frac{B_{j_1,j_2,j}}{2j+1}  = \frac{B_{j,j_2,j_1}}{2 j_1+1}  = \frac{B_{j_1,j,j_2}}{2 j_2+1}.
\end{equation}

We can exploit this equivalence by further restricting $j_2 \leq j_1 \leq j$, in which case the number of distinct bispectrum coefficients drops to $(J+1) (J + 2) (J + \frac{3}{2}) / 3$, a three-fold reduction in the limit of large $J$.

\subsection{SNAP Potential Energy Function}
\label{subsec:snap}

Given the bispectrum components as descriptors of the neighborhood of each atom, it remains to express the potential energy of a configuration of $N$ atoms in terms of these descriptors.  
We write the energy of the system containing $N$ atoms with positions $\textbf{r}^N$ as the sum of a reference energy $E_{ref}$ and a local energy $E_{local}$

\begin{equation}
E(\textbf{r}^N) = E_{ref}(\textbf{r}^N) + E_{local}(\textbf{r}^N).
\end{equation}

The reference energy includes known physical phenomena, such as long-range electrostatic interactions, for which well-established energy models exist.  $E_{local}$ must capture all the additional effects that are not accounted for by the reference energy.  Following Bart{\'{o}}k et al.~\cite{Bartok2010, Bartok2013} we assume that the local energy can be decomposed into separate contributions for each atom,

\begin{equation}
E_{local}(\textbf{r}^N) = \sum_{i=1}^{N} E_i (\textbf{q}_i)
\end{equation}

where $E_i$ is the local energy of atom $i$, which depends on the set of descriptors $\textbf{q}_i$, in our case the set of $K$ bispectrum components 
$\textbf{B}^i = \{ B_1^i, \ldots, B_K^i \} $.  The original GAP formulation of Bart{\'{o}}k et al.~\cite{Bartok2010} expressed the local energy in terms of a Gaussian process kernel. For the materials that we have examined so far, we have found that energies and forces obtained from quantum electronic structure calculations can be accurately reproduced by linear contributions from the lowest-order bispectrum components, with linear coefficients that depend only on the chemical identity of the central atom:

\begin{equation}
\label{eqn:snapatomenergy}
E^i_{SNAP}(\textbf{B}^i) = \beta^{\alpha_i}_0 + \sum_{k=1}^K \beta_k^{\alpha_i} B_k^i = \beta^{\alpha_i}_0 + \boldsymbol\beta^{\alpha_i}\cdot {\bf B}^i
\end{equation}

where $\alpha_i$ is the chemical identity of atom $i$ and $\beta_k^{\alpha}$ are the linear coefficients for atoms of type $\alpha$.  Hence the problem of generating the interatomic potential has been reduced to that of choosing the best values for the linear SNAP coefficients.  Since our goal is to reproduce the accuracy of quantum electronic structure calculations for materials under a range of conditions, it makes sense to select SNAP coefficients that accurately reproduce quantum calculations for small configurations of atoms representative of these conditions.  

In quantum methods such as density functional theory~\cite{Schultz2005} the most readily computed properties are total energy, atom forces, and stress tensor.  The linear form of the SNAP energy allows us to write all of these quantities explicitly as linear functions of the SNAP coefficients.  We restrict ourselves here to the case of atoms of a single type, but the results are easily extended to the general case of multiple atom types. In the case of total energy, the SNAP contribution can be written in terms of the bispectrum components of the atoms

\begin{equation}
\label{eqn:snapenergy}
E_{SNAP}(\textbf{r}^N) = N \beta_0 + \boldsymbol\beta\cdot\sum_{i=1}^{N} {\bf B}^i
\end{equation}

where $\boldsymbol\beta$ is the $K$-vector of SNAP coefficients and  $\beta_0$ is the constant energy contribution for each atom.  $ {\bf B}^i$ is the $K$-vector of bispectrum components for atom $i$.  The contribution of the SNAP energy to the force on atom $j$ can be written in terms of the derivatives of the bispectrum components w.r.t. ${\bf r}_j$, the position of atom $j$

\begin{equation}
\label{eqn:snapforce}
{\bf F}^j_{SNAP} = -\nabla_j E_{SNAP}  = - \boldsymbol\beta\cdot \sum_{i=1}^{N} \frac{\partial {\bf B}^i}{\partial {\bf r}_j},
\end{equation}

where ${\bf F}^j_{SNAP}$ is the force on atom $j$ due to the SNAP energy.  Finally, we can write the contribution of the SNAP energy to the stress tensor

\begin{equation}
\label{eqn:snapstress}
{\bf W}_{SNAP} = - \sum_{j=1}^{N}  {\bf r}_j \otimes \nabla_j E_{SNAP} =  - \boldsymbol\beta\cdot  \sum_{j=1}^{N}  {\bf r}_j \otimes \sum_{i=1}^{N} \frac{\partial {\bf B}^i}{\partial {\bf r}_j}
\end{equation}

where ${\bf W}_{SNAP}$ is the contribution of the SNAP energy to the stress tensor and $\otimes$ is the Cartesian outer product operator.  

All three of these expressions consist of the vector $\boldsymbol\beta$ of SNAP coefficients multiplying a vector of quantities that are calculated from the bispectrum components of atoms in a configuration.  This linear structure greatly simplifies the task of finding the best choice for $\boldsymbol\beta$.  We can define a system of linear equations whose solution corresponds to an optimal choice for $\boldsymbol\beta$, in that it minimizes the sum of square differences between the above expressions and the corresponding quantum results defined for a large number of different atomic configurations.  This is described in more detail in the following section.

\section{Automated Potential Generation Methodology} 
The previous section outlined the SNAP formulation.  
In practice, one needs to determine the values of the SNAP coefficients, $\boldsymbol\beta$. 
This section presents how we solve for the $K$-vector $\boldsymbol\beta$ of SNAP coefficients using a least-squares formulation.

\subsection{Formulation of the Linear Least Squares Problem}
\label{subsec:linear}
The fitting problem is overdetermined, in the sense that the number of data points that we are fitting to far exceeds the number of SNAP coefficients.   The cost of evaluating the bispectrum components $B_{j_1,j_2,j}$ increases strongly with the order of the indices $j$, $j_1$, and $j_2$. For this reason, $K$ is limited to the range $10 - 100$.  In contrast, with the availability of high-performance computers and highly optimized electronic structure codes, it is not difficult to generate data for hundreds or thousands of configurations of atoms.   Note that we refer to a \emph{configuration} as a set of atoms located at particular positions in a quantum mechanical calculation.  In most cases, the atoms define an infinite repeating structure with specified periodic lattice vectors.  For a particular configuration $s$, containing $N_s$ atoms, the electronic structure calculation yields $3 N_s + 7$ data values: the total energy, the $3 N_s$ force components, and the 6 independent components of the stress tensor.  The same quantities are calculated for the reference potential.  In addition, the bispectrum components and derivatives for each atom in the configuration are calculated.  We can use all of this data to construct the following set of linear equations.    

\begin{equation}
 \begin{bmatrix}
   \vdots  & \vdots  \\
   N_s  & \sum_{i=1}^{N_s}{\bf B}^i \\
   \vdots & \vdots \\
   0  & -\sum_{i=1}^{N_s}\frac{\partial {\bf B}^i}{\partial r_j^\alpha} 
  \\
   \vdots  & \vdots  \\
   0  & -\sum_{j=1}^{N_s}r_j^{\alpha}\sum_{i=1}^{N_s}\frac{\partial {\bf B}^i}{\partial r_j^\beta} & 
  \\
   \vdots  & \vdots  \\
  \end{bmatrix} \cdot
   \begin{bmatrix}
   \beta_0 \\
   \boldsymbol\beta\\
  \end{bmatrix} 
  = 
  \begin{bmatrix}
   \vdots \\
   E_s^{qm}-E_s^{ref}\\
   \vdots \\
   F^{qm}_{j,\alpha}-F^{ref}_{j,\alpha} \\
   \vdots\\
   W_{\alpha\beta,s}^{qm}-W_{\alpha\beta,s}^{ref} \\
   \vdots\\
  \end{bmatrix} 
\label{eqn:abmatrix}
\end{equation}

This matrix formulation is of the type ${\bf A}\cdot\boldsymbol\beta={\bf y}$ which can be solved for the coefficients $\boldsymbol\beta$.  
The optimal solution $\hat{\boldsymbol\beta}$ for this set of equations is~\cite{Strang1980}: 
\begin{equation} 
\hat{\boldsymbol\beta} = \underset{\boldsymbol\beta}{\operatorname{argmin}} \| ({\bf A}\cdot\boldsymbol\beta-{\bf y}) \|^2 ={\bf A}^{-1}\cdot{\bf y}
\end{equation}
In practice, we do not explicitly take the inverse of the ${\bf A}$ matrix, but instead use a QR factorization to solve for $\boldsymbol\beta$.  We have found the linear solve to obtain the optimal SNAP coefficients to be very fast and not poorly conditioned.  

We have also added the capability to perform weighted least squares to weight certain rows more than others.  That is, we add a vector of weights ${\bf w}$ to the minimization formulation:   

\begin{equation} 
\hat{\boldsymbol\beta} = \underset{\boldsymbol\beta}{\operatorname{argmin}} \| {\bf w} \circ ({\bf A} \cdot \boldsymbol\beta-{\bf y}) \|^2
\end{equation}

where $\circ$ is used to denote element by element multiplication by the weight vector.  Thus, each row in the ${\bf A}$ matrix and the ${\bf y}$ vector are multiplied by a weight specified for that row.  
In this way, we are able to specify weights per configuration type (e.g. BCC crystals, liquids, etc.) and per quantity of interest (e.g. energy, force, stress tensor).  We have found that the ability to weight different rows in the ${\bf A}$ matrix is critical to ensure the regression works well.  One reason is that the total energy, forces, and stress components can vary considerably in relative magnitude, depending on what units they are expressed in.  However, the more important reason is that it is desirable to control the relative influence of different configurations, depending on the material properties that are of greatest importance.  
For example, if we want the SNAP potential to more accurately reproduce BCC elastic constants, we can increase the weight on the stress components of strained BCC configurations.  We also found it helpful to convert all extensive quantities to intensive quantities, in order to counteract overweighting of configurations with large $N_s$.  Total energy rows were scaled by the number of atoms and stress tensor components rows were scaled by the cell volume. 

\begin{figure}[htbp]
  \centering
  \includegraphics[scale=0.4]{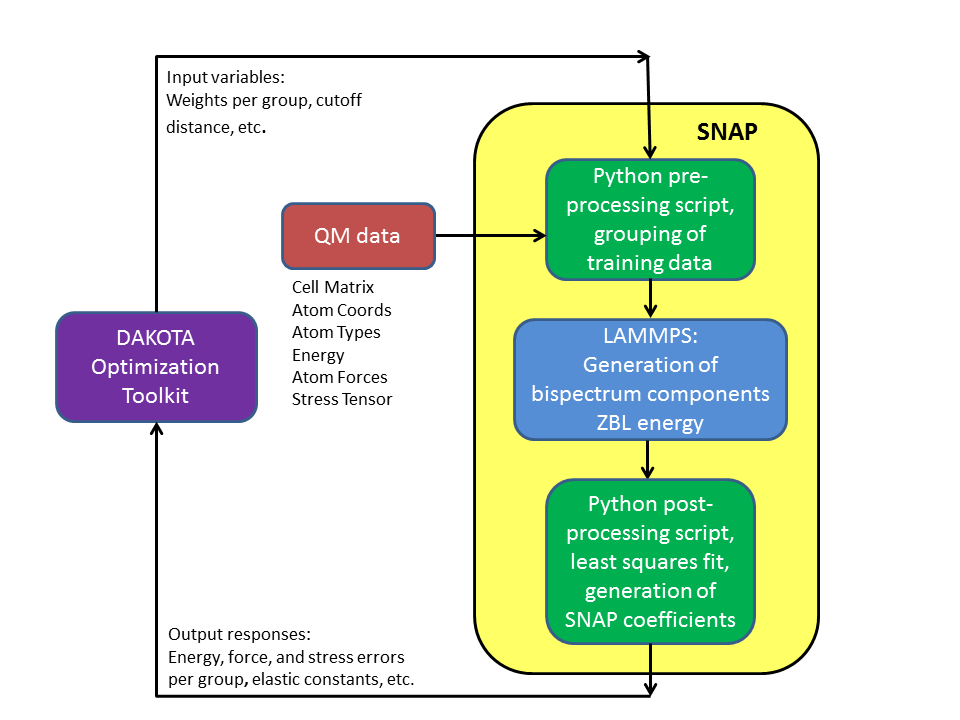}
  \caption{Flowchart of the optimization loop around the generation of the SNAP potential in LAMMPS}
  \label{fig:flowDakota}
\end{figure}
  
\subsection{SNAP software implementation}
\label{subsec:sw_implement}
We describe the Python framework to generate the SNAP fit within the LAMMPS software tool.  
We start with quantum mechanical (QM) training data, 
generated from ab initio calculations, which can be obtained from a wide variety of quantum electronic
structure software packages.
The training data is first converted to a set of files, one per configuration, 
using standardized JavaScript Object Notation (JSON) format.  The JSON files contain all relevant information 
such as atom coordinates, atom types, periodic cell dimensions, energies, forces, and stress tensors.
Note that  
the performance of the SNAP potential will depend on the comprehensiveness of the configurations 
in the training set.  In the example of tantalum below, we are interested in material plasticity and stability as well as elastic constants and the lattice parameter.  For this reason, the training data included configurations for all the important crystal structures, generalized stacking faults, free surfaces, liquid structures, and randomly deformed primitive cells of the ground-state
crystal. 

Once the training data is generated, pre-processing and post-processing scripts must be run to generate the SNAP potential in the LAMMPS software.
 The pre-processing script converts 
the training data from the JSON format into the native LAMMPS input format.  
LAMMPS is then used to 
generate the bispectrum components for the training data configurations, as well as calculating the reference potential.  The LAMMPS output is used to generate the 
energy, force and stress tensor rows of the $A$ matrix defined in Eq.~\ref{eqn:abmatrix}.  
The LAMMPS output for the reference potential is combined with the training data to generate the righthand side vector, which is the difference between the training data and 
the reference potential. A standard linear algebra
library is used to perform QR factorization, yielding the least-squares optimal SNAP linear coefficients, ${\hat{\boldsymbol\beta}}$.  The steps involved in the generation of the SNAP potential are shown in the yellow box on the right side of Fig.~\ref{fig:flowDakota}.

\subsection{Optimization of hyperparameters governing the SNAP potential}
Each SNAP potential developed in this manner depends on the particular values chosen for a set of ``hyperparameters.'' These include things such as the weights for each training configuration, the list of bispectrum components to be calculated, and the cutoff distance defining the neighborhood of an atom.  It is not obvious how to choose values of these hyperparameters which will lead to an ``optimal'' SNAP potential in the sense of minimizing the SNAP prediction error with respect to energy errors, force errors, or other quantities.   

To determine the optimal hyperparameters governing the SNAP potential, we 
have used an optimization framework and placed it around the SNAP calculation to generate many instances of SNAP potentials.
The optimization framework we use is the DAKOTA software~\cite{Dakota2011}, which is a toolkit of optimization and uncertainty quantification methods designed to interface to scientific computing codes.  The process of generating a SNAP potential within an optimization loop to optimize the governing parameters is shown in Fig.~\ref{fig:flowDakota}.  DAKOTA varies input parameters such as the weights per configuration group and the cutoff distance.  These values are then specified in the SNAP generation and the SNAP potential is calculated.  Once the SNAP coefficients are generated, they are used to predict the energy and forces of the QM training data.  The errors in the SNAP prediction (defined as the difference between SNAP configuration energies vs. QM configuration energies, SNAP forces vs. QM forces, etc.) are then aggregated into an objective function which is returned to DAKOTA and used as the quantity which DAKOTA tries to optimize. 

Using this optimization framework, we can identify the SNAP potential that has the ``best'' result, according to minimizing an objective function.  We have examined various objective functions.  Currently the objective function we use involves a weighted sum of the errors with respect to energies, forces, and stress tensors, as well as errors with respect to the elastic constants.  The objective function is not trivial to define:  the resulting SNAP potential can be sensitive to which error measures are weighted more in the objective function.  

\section{Implementation}
\label{sec:implementation}

A detailed account of the SNAP implementation and its optimization for specific computing platforms is given elsewhere~\cite{Trott2014}.
The force on each atom due to the SNAP potential is formally given by Eq.~\ref{eqn:snapforce}.  
In order to perform this calculation efficiently, we use a
neighbor list, as is standard practice in the LAMMPS code~\cite{LAMMPS,Plimpton1995}.
This list identifies
all the neighbors of a given atom $i$.   In order to avoid negative
and half-integer indices, we have switched notation from $u^j_{m,m'}$ to
$u^\eta_{\mu,\mu'}$, where $\eta=2 j$, $\mu = m+j$, and $\mu' = m'+j$. 
Analogous transformations are used for 
$\hcoeff{j}{m}{m'}{j_1}{\!m_1}{\!m'_1}{j_2}{m_2}{m'_2}$
and  $B_{j_1,j_2,j}$.
This also allows us to reclaim the symbol $j$ for indexing the neighbor atoms of atom $i$.  Finally, 
boldface symbols with omitted indices such as ${\bf u}_i$ are used to indicate a
finite multidimensional arrays of the corresponding indexed variables.

\begin{figure}[b!]
\begin{center}
\fbox{
\begin{minipage}{\textwidth}
\begin{tabbing}
xxx\=xxx\=xxx\=xxxxxxxxxxxxxxxxxxxxxxxxxxxxxxxxxxxxxxxxxx\= \kill
{\bf Compute\_SNAP():} \\
for $i$ in natoms() \{ \\
\> ${\bf u}_i$ = Calc\_U($i$) \\
\> ${\bf Z}_i$ = Calc\_Z($i$,${\bf u}_i$) \\
\> for $j$ in neighbors($i$) \{ \\
\> \> $\nabla_j {\bf u}_i$ = Calc\_dUdR($i, j$,${\bf u}_i$) \\
\> \> $\nabla_j {\bf B}^i$ = Calc\_dBdR($i, j$,${\bf u}_i$,${\bf Z}_i$,$\nabla_j {\bf u}_i$) \\
\> \> ${\bf F}_{ij} = - \boldsymbol\beta\cdot \nabla_j {\bf B}^i$ \\
\> \> ${\bf F}_i$ += $-{\bf F}_{ij} $; ${\bf F}_j$ += ${\bf F}_{ij} $ \\
\} \>\} 
\end{tabbing}
\end{minipage}
}
\end{center}

\caption{Base algorithm for the SNAP force calculation.}
\label{fig:calcsnap}
\vspace{-0.1cm}
\end{figure}
\begin{figure}[t!]
\begin{center}
\fbox{
\begin{minipage}{\textwidth} \begin{tabbing}
xxx\=xxx\=xxx\=xxx\=xxx\=xxx\=xxx\=xxxxxxxxxxxxxxxxxxxxxxx\= \kill
{\bf Function Calc\_dBdR($i, j$):} \\
for $(\eta, \eta_1, \eta_2)$ in GetBispectrumIndices() \{ \\
\> $\nabla_j B_{\eta_1,\eta_2,\eta} = 0 $\\
\> for ($\mu = 0; \mu \le \eta; \mu$++ ) \{\\
\> \> for ($\mu' = 0; \mu' \le \eta; \mu'$++) \{\\
\> \> \> $\nabla_j Z_{\eta_1,\eta_2,\eta}^{\mu,\mu'} = 0 $\\
\> \> \> for ($\mu_1 = \max{(0,\mu+(\eta_1-\eta_2-\eta)/2)}; $ \\ 
\> \> \> \> \> $\mu_1 \le \min{(\eta_1,\mu+(\eta_1+\eta_2-\eta)/2)}; \mu_1$++)  \{\\
\> \> \> \> for ($\mu_1' = \max{(0,\mu'+(\eta_1-\eta_2-\eta)/2)}; $ \\
\> \> \> \> \> \> $\mu_1' \le \min{(\eta_1,\mu'+(\eta_1+\eta_2-\eta)/2)}; \mu_1'$++) \{\\
\> \> \> \>  \> $\mu_2 = \mu-\mu_1$ ; $\mu'_2 = \mu'-\mu'_1$\\
\> \> \> \>  \> $\nabla_j Z_{\eta_1,\eta_2,\eta}^{\mu,\mu'} $ += $ H_{\eta_1,\mu_1,\mu'_1,\eta_2,\mu_2,\mu'_2}^{\eta,\mu,\mu'}$ \\
\> \> \> \>  \> \> $ (u^{\eta_1}_{\mu_1,\mu'_1} \nabla_j u^{\eta_2}_{\mu_2,\mu'_2} + u^{\eta_2}_{\mu_2,\mu'_2} \nabla_j u^{\eta_1}_{\mu_1,\mu'_1} )$\\
\> \> \>\} \> \} \\
\> \> \> $\nabla_j B_{\eta_1,\eta_2,\eta} $ += $(u^{\eta}_{\mu,\mu'})^* \nabla_j Z_{\eta_1,\eta_2,\eta}^{\mu,\mu'} +  Z_{\eta_1,\eta_2,\eta}^{\mu,\mu'} (\nabla_j u^{\eta}_{\mu,\mu'})^*$\\
\}\>\}\>\}
\end{tabbing}
\end{minipage}
}
\end{center}
\caption{Original algorithm for the derivative of the bispectrum components of atom $i$ w.r.t. the position of atom $j$ using Eq.~\ref{eqn:dbisold}. }
\label{fig:calcdbidrjold}
\vspace{-0.1cm}
\end{figure}

\begin{figure}[t!]
\begin{center}
\fbox{
\begin{minipage}{\textwidth} \begin{tabbing}
xxx\=xxx\=xxx\=xxx\=xxx\=xxx\=xxx\=xxxxxxxxxxxxxxxxxxxxxxx\= \kill

{\bf Function Calc\_dBdR($i, j$):} \\
for $(\eta, \eta_1, \eta_2)$ in GetBispectrumIndices() \{  \\
\> $\nabla_j B_{\eta_1,\eta_2,\eta} = 0 $\\
\> for ($\mu = 0; \mu \le \eta; \mu$++ ) \{ \\
\> \> for ($\mu' = 0; \mu' \le \eta; \mu'$++) \{  \\
\> \> \> $\nabla_j B_{\eta_1,\eta_2,\eta} $ += $Z_{\eta_1,\eta_2,\eta}^{\mu,\mu'} (\nabla_j u^{\eta}_{\mu,\mu'})^*  $\\
\>\} \> \} \\
\> for ($\mu_1 = 0; \mu_1 \le \eta_1; \mu_1$++ ) \{ \\
\> \> for ($\mu'_1 = 0; \mu'_1 \le \eta_1; \mu'_1$++) \{  \\
\> \> \> $\nabla_j B_{\eta_1,\eta_2,\eta} $ += $\frac{\eta+1}{\eta_1+1} Z_{\eta,\eta_2,\eta_1}^{\mu_1,\mu'_1} (\nabla_j u^{\eta_1}_{\mu_1,\mu'_1})^* $\\
\>\} \> \} \\
\> for ($\mu_2 = 0; \mu_2 \le \eta_2; \mu_2$++ ) \{ \\
\> \> for ($\mu'_2 = 0; \mu'_2 \le \eta_2; \mu'_2$++) \{ \\
\> \> \> $\nabla_j B_{\eta_1,\eta_2,\eta} $ += $\frac{\eta+1}{\eta_2+1}  Z_{\eta_1,\eta,\eta_2}^{\mu_2,\mu'_2} (\nabla_j u^{\eta_2}_{\mu_2,\mu'_2})^*$\\
\>\} \> \} \\
\}
\end{tabbing}
\end{minipage}
}
\end{center}

\caption{Improved algorithm for the derivative of the bispectrum components of atom $i$ w.r.t. the position of atom $j$ using Eq.~\ref{eqn:dbisnew}.}
\label{fig:calcdbidrjnew}
\vspace{-0.1cm}
\end{figure}

Fig.~\ref{fig:calcsnap} gives the resulting force computation algorithm, where 
Calc\_U() calculates all expansion coefficients $u^\eta_{\mu,\mu'}$ for an atom $i$ 
while Calc\_Z() calculates the partial sums $Z_{\eta_1,\eta_2,\eta}^{\mu,\mu'} $,
which are defined as

\begin{equation}
\label{eqn:Z}
Z_{j_1,j_2,j}^{m,m'}  = \\
\sum_{m_1,m'_1=-j_1}^{j_1}\sum_{m_2,m'_2=-j_2}^{j_2}
H_{j_1,m_1,m'_1,j_2,m_2,m'_2}^{j,m,m'} u^{j_1}_{m_1,m'_1} u^{j_2}_{m_2,m'_2} .
\end{equation}

In the loop over neighbors, first the derivatives of ${\bf u}_i$ with respect to the distance vector between atoms $i$ and $j$ 
are computed in Calc\_dUdR() and then the derivatives of $B^i$ are computed in Calc\_dBdR(), which is the most computationally expensive 
part of the algorithm.    For the parameter sets used in this study, 
Calc\_dBdR() is responsible for approximately 90\% of all floating point and memory operations. Thus, we concentrate our description on this function.
In Fig.~\ref{fig:calcdbidrjold} we show the original algorithm based on Eq.~\ref{eqn:dbisold}.  In Fig.~\ref{fig:calcdbidrjnew} we show the improved algorithm 
based on Eq.~\ref{eqn:dbisnew} that takes advantage of the symmetry relation Eq.~\ref{eqn:symm}.  The elimination of the two innermost nested loops
reduces the scaling of the computational complexity of the bispectrum component from $O(J^4)$ to $O(J^2)$ where $J$ is the upper limit on $j_1$, $j_2$, and $j$.  

\subsection{Scaling Results}
\label{subsec:scaling}

\begin{figure}[t!]
  \centering
  \includegraphics[scale=0.4]{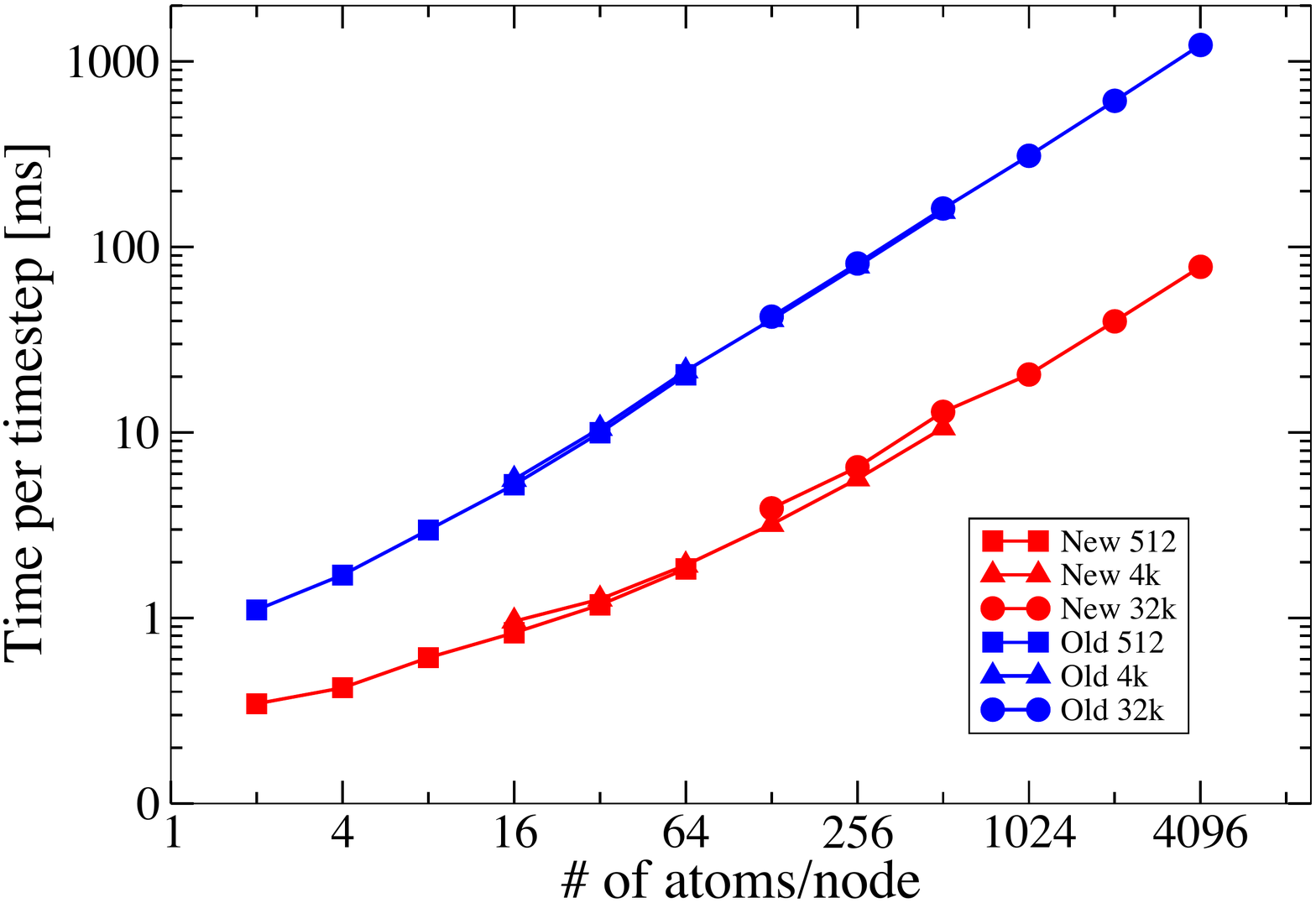}
  \caption{CPU time per MD time step versus atoms per node for benchmark simulations of BCC tantalum using the old and improved implementations of the SNAP potential.  
  Results are shown for systems containing 512, 4096, and 32768 atoms. The number of nodes ranged from 8 to 256 nodes.}

  \label{fig:timing}
\end{figure}

\begin{figure}[t!]
  \centering
  \includegraphics[scale=0.4]{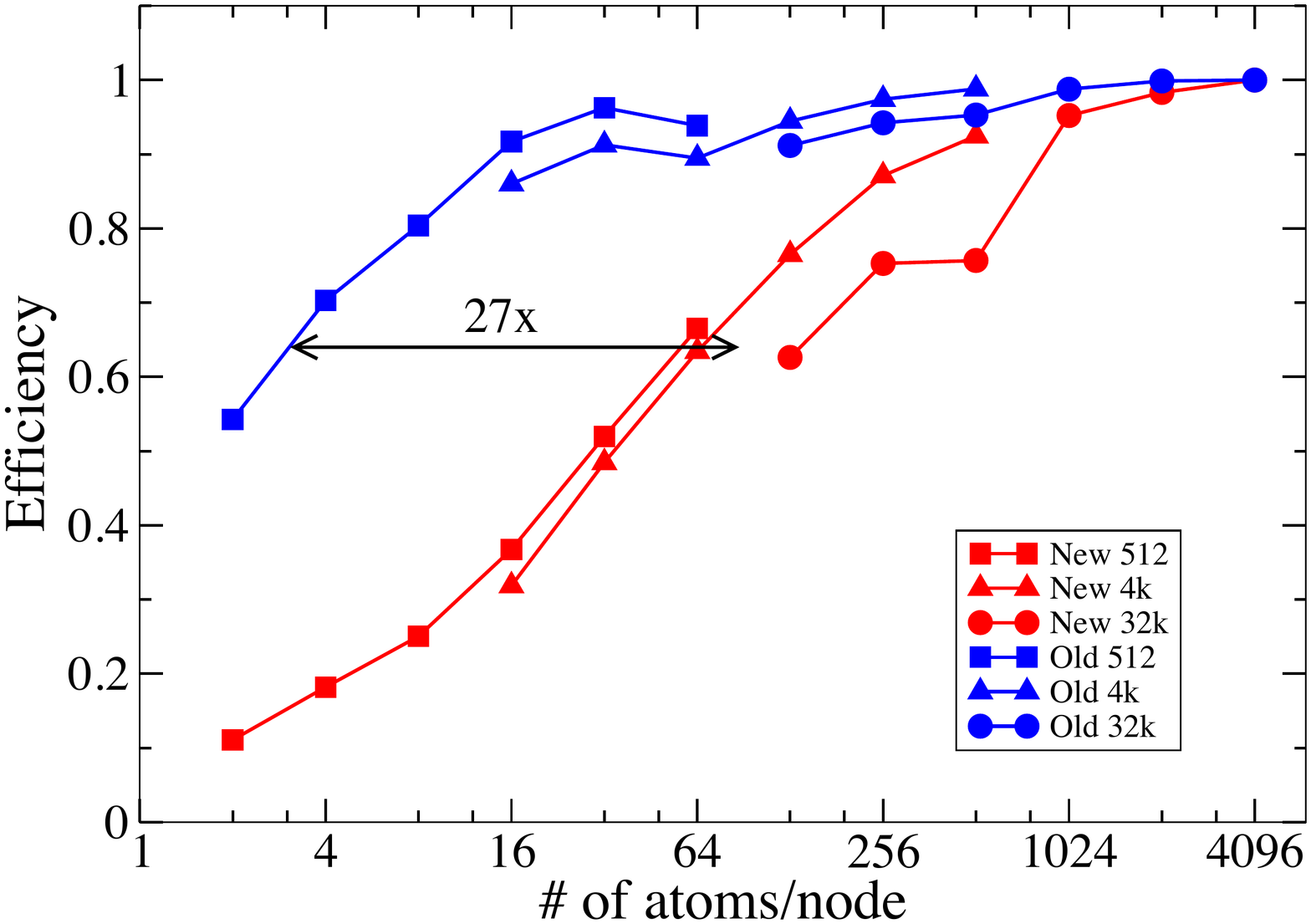}
  \caption{Parallel efficiency versus atoms per node for benchmark simulations of BCC tantalum using the old and improved implementations of the SNAP potential.  
  Results are shown for systems containing 512, 4096, and 32768 atoms. The number of nodes ranged from 8 to 256 nodes.}

  \label{fig:efficiency}
\end{figure}

Both of these algorithms have been implemented in the LAMMPS parallel molecular dynamics package~\cite{LAMMPS,Plimpton1995}.  
In Fig.~\ref{fig:timing} we compare the performance of the old and new algorithms.  Timings are based on a 10,000 step MD simulations 
of BCC tantalum crystal using the SNAP potential described in Section \ref{sec:tantalum}.  The calculations were performed on Sandia's Chama 
high-performance cluster with a dual socket Intel Sandy Bridge processor with 16 cores on each node.  Three different system sizes were used,
containing 512, 4096, and 32768 atoms.  Each system size was run on 8, 16, 32, 64, 128, and 256 nodes. We plot the time required to calculate one MD time step 
versus the number of atoms per node.  In this form, results for the three different system sizes are almost indistinguishable, indicating that both single node
performance and strong scaling efficiency are determined primarily by the number of atoms per node.  When the number of atoms per node is large, the
parallel scaling is close to ideal, and the improved algorithm is consistently about 16x faster than the old algorithm.  As the number of atoms per node
decreases, this margin also decreases, due to the parallel efficiency of the new algorithm decreasing more rapidly.  This is seen more clearly in 
Fig.~\ref{fig:efficiency} which shows the parallel efficiency of each calculation relative to the timing for the largest system on 8 nodes.  Both
the old and new algorithms show a decrease in parallel efficiency as the number of atoms per node decreases, but this happens at an earlier point
for the new algorithm, because there is a lot less computation per atom, while the amount of communication is unchanged.

\section{SNAP Potential for Tantalum}
\label{sec:tantalum}

\subsection{Training Data}
\label{subsec:trainingdata}

The training set data as well as the validation data for these potentials were computed using density functional theory (DFT) electronic structure calculations as implemented in the Vienna Ab initio Simulation Package (VASP)~\cite{Kresse1996}.  The pseudopotentials employed are of the projector augmented wave (PAW) form~\cite{Kresse1999} with the exchange correlation (XC) energy evaluated using the Perdew, Burke and Ernzerhof (PBE) formulation~\cite{Perdew1996} of the generalized gradient approximation (GGA).  The pseudopotential employed treats 11~electrons $(5p5d6s)$ as valent.  Compared to the use of pseudopotentials with 5 valence states, this improved the prediction of the elastic constants and is expected to be important in geometries with highly compressed bond lengths.  A plane-wave cut-off energy of 500~eV was used to insure the convergence of the computed stresses and forces in addition to the energies.  The k-space integrations were performed using the Monkhorst-Pack (MK) integration scheme~\cite{Monkhorst1976} with mesh sizes that depended on the particular geometry and chosen to obtain convergence of the energy, forces and stress tensor.  The lattice predicted for Ta is 0.332~nm compared to the experimental value of 0.331~nm.  The computed elastic constants, $C_{11}$, $C_{12}$ and $C_{44}$ are 267.5, 159.7, and 71.1~GPa.  This compares well with the experimental results of 266.3, 158.2, and 87.4~GPa by Featherston and Neighbours~\cite{Featherston1963} and 266, 161.1, and 82.5~GPa by Katahara et al.~\cite{Katahara1976}.

A variety of different types of atomic configurations were used to construct the full set of training data, and these are summarized in Table~\ref{tab:trainingdata}.  Configurations of different types were chosen to adequately sample the important regions of the potential energy surface.  Configurations of type ``Displaced'' were constructed by randomly displacing atoms from their equilibrium lattice sites in supercells of the A15, BCC, and FCC crystal structures.  The maximum displacement in any direction was limited to 10\% of the nearest neighbor distance. Configurations of type``Elastic'' were constructed by applying random strains to primitive cells of the BCC crystal.  The maximum linear strains were limited to 0.5\%.  The configurations of type ``GSF" consist of both relaxed and unrelaxed generalized stacking faults on the (110) and (112) crystallographic planes in the $\langle111\rangle$ direction. The configurations of type ``Liquid" were taken from a high-temperature quantum molecular dynamics simulations of molten tantalum.  The configurations of type ``Surface" consisted of relaxed and unrelaxed (100), (110), (111), and (112) BCC surfaces. For each type of configuration we specified a weight for the energy, force, and stress.  We set the force and energy weights  of the``Elastic" configurations to zero and we set the stress weights of all other configurations to zero.

\begin{table}[t]
\begin{center}
\begin{tabular}{llllll}
\hline\hline 
Type & $N_{conf}$  & $N_{atoms}$ & Energy & Force & Stress  \\ 
\hline
Displaced A15   & 9 & 64 & 100 & 1 & - \\        
Displaced BCC  & 9 & 54 & 100 & 1 & -  \\
Displaced FCC  & 9 & 48 & 100 &1 & - \\
Elastic BCC       &100 & 2 & - & - & 0.0001 \\
GSF 110            & 22 & 24 & 100 & 1 & - \\
GSF 112            & 22 & 30 & 100 & 1 & - \\
Liquid                  & 3 & 100 & 100 & 1 & - \\
Surface               & 7 & 30 & 100 & 1 & - \\
\hline\hline
\end{tabular}
\end{center}
\caption{DFT data used to fit the SNAP potential for Tantalum}
\label{tab:trainingdata}
\end{table}

In addition to the training data and way in which different quantities were weighted, the quality of the SNAP potential also was somewhat dependent on the choices made for the reference potential and the SNAP hyperparameters.  Because the training data did not sample highly compressed configurations, it was important that the reference potential provide a good physical description of Pauli repulsion that dominates the interaction at close separation.  We chose the Ziegler-Biersack-Littmark (ZBL) empirical potential that has been found to correctly correlate the high-energy scattering of ions with their nuclear charge $Z_{zbl}$~\cite{Ziegler1985}.  Because the ZBL potential decays rapidly with radial separation, we used a switching function to make the energy and force go smoothly to zero at a distance $R_{zbl,o}$, while leaving the potential unchanged for distances less than $R_{zbl,i}$.  The values for these three parameters are given in Table~\ref{tab:snapparams}.   For the neighbor density switching function, we use the same functional form as Bart{\'{o}}k et al.~\cite{Bartok2010}.

\begin{eqnarray}
\label{eqn:f_c}
f_c(r)  & = & \frac{1}{2}(\cos(\pi r/R_{cut}) + 1), r \leq R_{cut} \\
& = & 0,  r > R_{cut}
\end{eqnarray}

The DAKOTA package was used to optimize the value of $R_{cut}$ so as to minimize the error in the energies, forces, and elastic constants relative to the training data.  The resultant value of $R_{cut} = 4.67637$ is physically reasonable, as it includes the 14 nearest neighbors in the BCC crystal, and the first coordination shell in the melt.  The effect of using fewer or more bispectrum components was examined experimentally by varying $J$.  We found that the fitting errors decreased monotonically with increasing $J$, but the marginal improvement also decreased. We found that truncating at $J = 3$ provided a good trade off between accuracy and computational efficiency.  The full set of ZBL and SNAP parameters values are given in Table~\ref{tab:snapparams}, while the values of the SNAP linear coefficients corresponding to each bispectrum component are listed in Table~\ref{tab:snapcoefficients}.~\cite{LAMMPS}

\begin{table}[t]
\begin{center}
\begin{tabular}{ll}
\hline\hline 
$J$  & 3 \\
$R_{cut}$ & 4.67637 \AA \\
$\theta_0^{max}$  & $0.99363\pi$ \\
$R_{zbl,i}$ & 4.0 \AA \\
$R_{zbl,o}$ & 4.8 \AA \\
$Z_{zbl}$ & 73.0 \\
\hline\hline
\end{tabular}
\end{center}
\caption{SNAP and ZBL potential parameters used to model tantalum.~\cite{LAMMPS}}
\label{tab:snapparams}
\end{table}

\begin{table}
\begin{center}
\begin{tabular}{llllr}
\hline\hline 
 $k$ & $2 j_1$  & $2 j_2$ & $2 j$ & $\beta_k$  \\ 
\hline
0  &     &    &    & -2.92477 \\  
1  &  0 & 0 & 0 & -0.01137 \\  
2  &  1 & 0 & 1 & -0.00775 \\  
3  &  1 & 1 & 2 & -0.04907 \\  
4  &  2 & 0 & 2 & -0.15047 \\  
5  &  2 & 1 & 3 & 0.09157 \\  
6  &  2 & 2 & 2 & 0.05590 \\  
7  &  2 & 2 & 4 & 0.05785 \\  
8  &  3 & 0 & 3 & -0.11615 \\  
9  &  3 & 1 & 4 & -0.17122 \\  
10 &  3 & 2 & 3 & -0.10583 \\  
11 &  3 & 2 & 5 & 0.03941 \\  
12 &  3 & 3 & 4 & -0.11284 \\  
13 &  3 & 3 & 6 & 0.03939 \\  
14 &  4 & 0 & 4 & -0.07331 \\  
15 &  4 & 1 & 5 & -0.06582 \\  
16 &  4 & 2 & 4 & -0.09341 \\  
17 &  4 & 2 & 6 & -0.10587 \\  
18 &  4 & 3 & 5 & -0.15497 \\  
19 &  4 & 4 & 4 & 0.04820 \\  
20 &  4 & 4 & 6 & 0.00205 \\  
21 &  5 & 0 & 5 & 0.00060 \\  
22 &  5 & 1 & 6 & -0.04898 \\  
23 &  5 & 2 & 5 & -0.05084 \\  
24 &  5 & 3 & 6 & -0.03371 \\  
25 &  5 & 4 & 5 & -0.01441 \\  
26 &  5 & 5 & 6 & -0.01501 \\  
27 &  6 & 0 & 6 & -0.00599 \\  
28 &  6 & 2 & 6 & -0.06373 \\  
29 &  6 & 4 & 6 & 0.03965 \\  
30 &  6 & 6 & 6 & 0.01072 \\  
\hline\hline
\end{tabular}
\end{center}
\caption{SNAP linear coefficients for tantalum.~\cite{LAMMPS}}
\label{tab:snapcoefficients}
\end{table}

\subsection{Validation Results}
\label{subsec:validationresults}

\begin{figure}[t]
\includegraphics[width=0.95\textwidth, clip=]{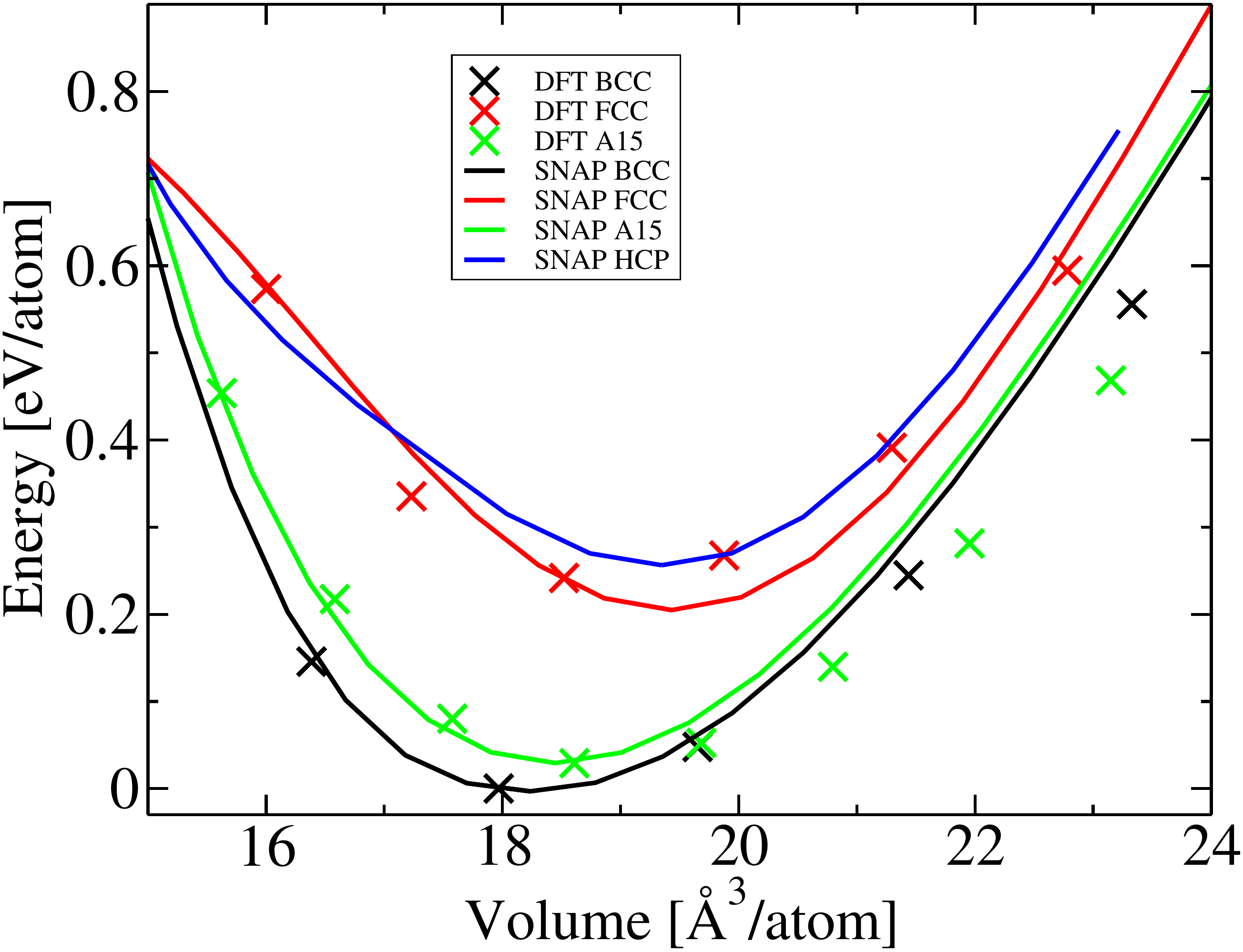}
\caption{Energy versus volume for various crystal structures as indicated in the legend, as computed by SNAP (solid curves) and from DFT (x).}
\label{fig:evsv} 
\end{figure}

One of the crucial requirements for interatomic potential models is that they predict the correct minimum energy crystal structure and that the energetics of competing crystal structures be qualitatively correct.  Fig.~\ref{fig:evsv} plots the energy per atom computed with the SNAP potential as a function of volume for the BCC, FCC, A15, and HCP phases.  The energy of diamond structure Ta was also computed.  As expected, it was found to be substantially ($\sim$2.8 eV/atom) higher than the BCC phase and is not included on the plot.  In addition, energies computed from density functional theory are included as crosses.  It is seen that the relative energy of these phases is correctly predicted.  The BCC phase is the most stable 
throughout the volume range considered, with the A15 phase somewhat higher.  The FCC phase is next with a minimum energy about 0.2 eV/atom above that of BCC.  Note that these energy differences are very consistent with the DFT calculations.  The SNAP predicted HCP energy is also shown.  Note that the SNAP potential is able to differentiate HCP and FCC crystal structures which are structurally very similar.  The SNAP potential predicts that the HCP structure is higher in energy than the FCC structure.  This is a prediction in that no HCP data was used in the potential construction. The relative energies of HCP and FCC are in agreement with our DFT calculations, which show that lowest energy HCP structure (not shown) lies 0.04 eV/atom above the minimum energy FCC structure.  Further, the SNAP potential predicts the HCP $c/a$ ratio to be 1.72, which is considerably greater than ideal value $c/a \approx 1.63$.  The DFT calculations for HCP (not shown) predict an even larger value of $c/a$ = 1.77.

\subsection{Melting point and Liquid structure}

The melting point predicted by the SNAP potential has been determined.  An atomistic slab was created and brought to temperature above the melting point using a Langevin thermostat until the surface of the slab was melted.  The molecular dynamics simulation was then continued in an NVE ensemble.  This resulted in a system containing two solid-liquid interfaces.  The temperature of the MD system now fluctuated around the equilibrium melting temperature.  It is important that the solid phase be at the correct melting point density.  This was ensured through a simple iterative procedure.  An estimate of the melting point was obtained for an assumed lattice constant, the lattice constant of the solid at that temperature was determined from an NPT simulation of the solid, and the melting point was determined with the interfacial area determined by this lattice constant.  This process was iterated until the assumed and predicted melting points agreed.  
This procedure predicted a melting point of 2790~K.   This value is in reasonable agreement with the experimental melting point of 3293~K.  It should be noted that the melting point is typically a difficult quantity for interatomic potentials to predict accurately.  Further, the comparison with experiment is not a direct test of the agreement of the SNAP potential with DFT calculations.  The experimental value reflects contributions of the free energy of electronic excitations.  Further, the DFT prediction for the melting point is not known.

\begin{figure}[t]
\includegraphics[width=0.95\textwidth, clip=]{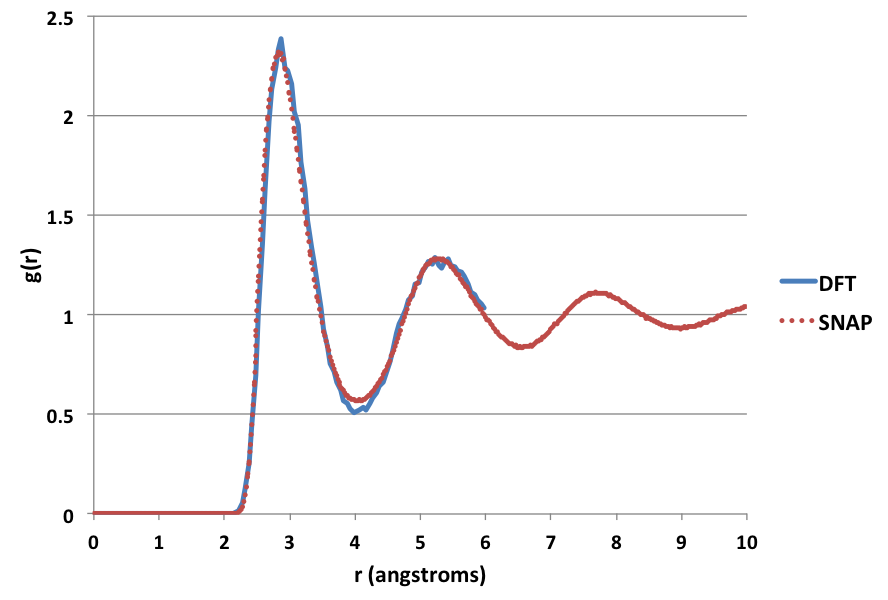}
\caption{Comparison of pair correlation function for molten tantalum calculated using DFT (blue) and SNAP (red)}
\label{fig:liquidstructure} 
\end{figure}

While configurations that correspond to molten Ta were used in the training set, it is important to determine if the potentials actually reproduce the correct distribution of spatial density correlations in the liquid state.  The liquid is an important test of potential models since it samples configurations that are far from those of the equilibrium solid crystals.  In particular, the liquid structure depends strongly on the repulsive interactions that occur when two atoms approach each other.  We calculated the pair correlation function of the liquid, $g(r)$, both using DFT and from the SNAP potential. These simulations were performed for the same temperature, 3250~K, and atomic density of 49.02~atom/nm$^3$.  The DFT simulation treated 100~atoms for a period of 2~ps while the SNAP simulations considered a cell containing 1024~atoms and averaged over 500 ps.  Fig.~\ref{fig:liquidstructure} compares $g(r)$ obtained in the two simulations.  The agreement is excellent except perhaps in the region of the first minimum.  Note that there is substantially more statistical uncertainty in the DFT result due to the short simulation time and the DFT structure can not be determined beyond about 0.6~nm, due to the smaller simulation cell.  These results indicate that the SNAP potential provides a good representation of the molten structure. 

\subsection{Planar and Point Defects}

A key metric for the applicability of an interatomic potential model is its ability to describe common defects in a material.   Table~\ref{tab:defectenergies} presents a comparison of the 
energy associated with surfaces, unstable stacking faults, vacancies and self-interstitial atoms.  The results obtained from the SNAP potential are compared to our DFT calculations and also against two other interatomic potential models, the embedded atom method (EAM) model developed by Zhou et al.~\cite{Zhou2004} and the angular dependent potential (ADP) due to Mishin and Lozovoi~\cite{Mishin2006}.

The training data includes the surfaces and unstable stacking faults presented in Table~\ref{tab:defectenergies}.  The unstable stacking fault energy is the maximum of the energy associated with translating half of the crystal in order to create a stacking fault on the specified plane.  This energy is expected to be related to the structure and behavior of dislocations.  
The overall magnitude of the surface energies is in good agreement with the DFT data.  The SNAP potential correctly predicts that the (110) surface is the lowest energy surface plane though it does not correctly predict the relative energies of the other higher energy surface planes. 
The surface energies predicted by the SNAP potential are generally somewhat better than those predicted by the EAM and ADP potentials.  
All of the potentials predict unstable stacking fault energies less than the DFT results though all of the potentials predict that the unstable stacking fault energy on the (112) plane is higher than on the (110) plane.

The point defect energies were not included in the training data for the SNAP potential.  The predicted vacancy formation energy is 0.15~eV lower than the density function value.  The EAM and ADP potentials are in better agreement with the DFT value.  However, both of those potentials were also explicitly fit to the vacancy formation energy.  The formation energy for self-interstitials are 
presented for four different configurations.  Self-interstitial energies and preferred geometries are challenging quantities for interatomic potential models since they typically involve bond lengths which are substantially shorter than equilibrium values.  The overall values for the SNAP potential are in reasonable accord with the DFT, but they incorrectly indicate that the $\langle$110$\rangle$ dumbbell is more favorable than the Crowdion configuration predicted by the DFT calculations. 

\begin{table}[t]
\begin{center}
\begin{tabular}{lllll}
\hline\hline 
 & DFT & SNAP & EAM & ADP \\ \hline
Lattice Constant (\AA)	& 3.320 & 3.316 & 3.303 & 3.305 \\
$B$ (Mbar) & 1.954 & 1.908 & 1.928 & 1.971 \\
$C' = \frac{1}{2}(C_{11} - C_{12})$ (Mbar) & 50.7 & 59.6 & 53.3 & 51.0 \\
$C_{44}$ (Mbar) & 75.3 & 73.4 & 81.4 & 84.6 \\
Vacancy Formation Energy (eV) & 2.89 & 2.74 & 2.97 & 2.92 \\
(100) Surface Energy (J/m$^2$) & 2.40 & 2.68 & 2.34 & 2.24 \\
(110) Surface Energy (J/m$^2$) & 2.25 & 2.34 & 1.98 & 2.13 \\
(111) Surface Energy (J/m$^2$) & 2.58 & 2.66 & 2.56 & 2.57 \\
(112) Surface Energy (J/m$^2$) & 2.49 & 2.60 & 2.36 & 2.46 \\
(110) Relaxed Unstable SFE (J/m$^2$) & 0.72 & 1.14 & 0.75 & 0.58 \\
(112) Relaxed Unstable SFE (J/m$^2$) & 0.84 & 1.25 & 0.87 & 0.74 \\
Self-Interstitial $-$ Octahedral Site (eV) & 6.01 & 7.10 & 5.06 & 7.61 \\
Self-Interstitial $-$ Crowdion (eV) & 4.73 & 5.74 & 5.09 & 7.02 \\
Self-Interstitial  $-$ $\langle 100 \rangle$ dumbbell (eV) & 6.12 & 6.89 & 5.24 & 7.59 \\
Self-Interstitial  $-$ $\langle$110$\rangle$ dumbbell (eV) & 5.63 & 5.43 & 4.93 & 6.99 \\\hline\hline
\end{tabular}
\end{center}
\caption{Calculated lattice constant, elastic constants, and formation energies for various crystal surfaces and defects using the SNAP, EAM~\cite{Zhou2004} , and ADP~\cite{Mishin2006} potentials, compared to DFT calculations.}
\label{tab:defectenergies}
\end{table}

\subsection{Dislocations}

\begin{figure}[t]
\includegraphics[width=0.95\textwidth]{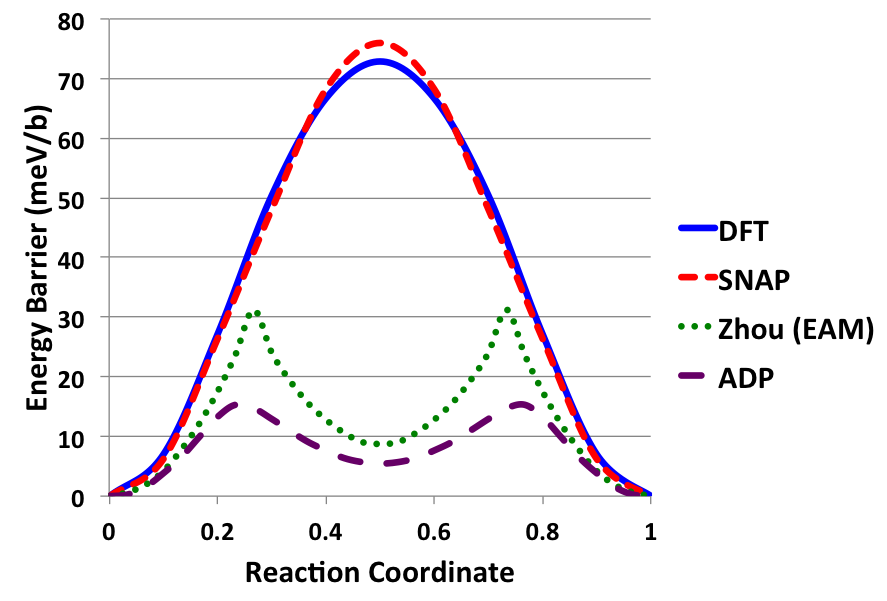}
\caption{Comparison of screw dislocation migration energy barrier calculated using DFT and the SNAP, EAM~\cite{Zhou2004}, and ADP~\cite{Mishin2006} potentials.}
\label{fig:dislocationbarrier} 
\end{figure}

Fig.~\ref{fig:dislocationbarrier} shows the screw dislocation migration energy barrier calculated using DFT and the SNAP, EAM~\cite{Zhou2004} , and ADP~\cite{Mishin2006} potentials.
One of the dominant deformation mechanisms for metallic materials is the motion of dislocations.  For the case of body-centered cubic materials such as tantalum, the screw dislocations are known to play a crucial role.
The structure and motion of screw dislocations in BCC metals has been examined for many years and is discussed in detail by Gr\"{o}ger et al.~\cite{Groger2008} and by references therein.  
A crucial feature of screw dislocations in BCC metals is the Peierls barrier which is the energy barrier to move the dislocation to its next stable configuration.  
Unlike face-centered-cubic metals where the Peierls barrier is generally negligible, the barrier in the case of BCC metals is substantial and plays a significant role in mechanical deformation.  
As has been shown recently by Weinberger et al.~\cite{Weinberger2013}, many empirical potentials for BCC metals predict qualitatively incorrect Peierls barriers.  
DFT calculations show that the Peierls has a simple shape with a single hump while many empirical potentials predict a transition path with two maxima and a metastable intermediate state.  
The Peierls barrier computed via the SNAP potential is shown in Fig.~\ref{fig:dislocationbarrier} along with the DFT barrier and the barriers predicted by the EAM and ADP potentials.  In all cases the barriers were computed based on a dislocation dipole configuration as described by Weinberger et al.~\cite{Weinberger2013}.  While the ADP and EAM potentials both predict incorrect barriers with an intermediate metastable state, the SNAP potential predicts a barrier with a single maximum in agreement with the DFT results.  Further, the magnitude of the barrier is in excellent agreement with the DFT prediction.    

\section{Summary}
\label{sec:summary}

In this paper we have introduced a new class of interatomic potentials based on the 4D bispectrum components
first proposed by Bart{\'{o}}k et al.~\cite{Bartok2010}.  Our SNAP potentials differ from Bart{\'{o}}k's GAP potentials primarily
in the use of an explicit linear dependence of the energy on the bispectrum components. 
We have developed a powerful machine-learning methodology for fitting SNAP potentials to large data sets of high-accuracy
quantum electronic structure calculations.  We have demonstrated the effectiveness of this approach for tantalum.
The new SNAP potential accurately represents a wide range of energetic properties of solid tantalum phases,
as well as both the structure of molten tantalum and its melting point. Most importantly,
we have found that the SNAP potential for tantalum correctly predicts the size and shape of the Peierls barrier
for screw dislocation motion in BCC tantalum.  This is a critical property for describing plasticity in tantalum
 under shear loading and is not correctly described by other published potentials for tantalum~\cite{Zhou2004,Mishin2006}.

One possible drawback of the SNAP methodology is computational cost when used in large-scale atomistic simulations. Even with
the algorithmic improvements described in this paper, the cost of SNAP is one to two orders of magnitude more
expensive than its competitors.  However, it is important to note the high computational intensity also makes
the method very amenable to massively parallel algorithms.  For a fixed size problem, the SNAP potential has been shown
to scale to a much greater extent than simpler potentials~\cite{Trott2014}.  We anticipate that atomistic simulations
of materials behavior using high-accuracy computationally-intensive potentials such as SNAP will become more
common as access to petascale computing resources increases.

\section*{Acknowledgement} 
The authors acknowledge helpful discussions with Stan Moore and Jonathan Moussa on the symmetry properties of bispectrum components.  Sandia National Laboratories is a multi-program laboratory managed and operated by Sandia Corporation, a wholly owned subsidiary of Lockheed Martin Corporation, for the U.S. Department of Energy's National Nuclear Security Administration under contract DE-AC04-94AL85000.




\nocite{*}
\bibliographystyle{model1-num-names}
\bibliography{snap_jcp}







\appendix

\section{Proof of additional symmetry relation on bispectrum components}
\label{sec:bispectrumsymmetry}

The bispectrum components for an arbitrary function defined on the 3-sphere are given by a sum over triple products of the expansion coefficients

\begin{equation}
\label{eqn:bispectrum}
B_{j_1,j_2,j}  = \\
\sum_{m,m'}^{j} (u^j_{m,m'})^*\sum_{m_1,m'_1}^{j_1}\sum_{m_2,m'_2}^{j_2} 
\hcoeff{j}{m}{m'}{j_1}{\!m_1}{\!m'_1}{j_2}{m_2}{m'_2}
u^{j_1}_{m_1,m'_1} u^{j_2}_{m_2,m'_2}
\end{equation}

where $u^{j}_{m,m'}$ are expansion coefficients, given by the inner product with the corresponding basis function $U^{j}_{m,m'}$.  The constants 
$\hcoeff{j}{m}{m'}{j_1}{\!m_1}{\!m'_1}{j_2}{m_2}{m'_2}$
are the Clebsch-Gordan coupling coefficients that can be written as products of the more well-known Clebsch-Gordan coefficients for functions on the 2-sphere

\begin{equation}
\hcoeff{j}{m}{m'}{j_1}{\!m_1}{\!m'_1}{j_2}{m_2}{m'_2} = C_{j_1,m_1,j_2,m_2}^{j,m} C_{j_1,m'_1,j_2,m'_2}^{j,m'}
\end{equation}

It follows from several elementary symmetry properties of the Clebsch-Gordan coefficients~\cite{Varshalovich1987} that coupling coefficients for which the indices $j_2$ and $j$ are interchanged satisfy the following identity

\begin{equation}
\frac{\hcoeff{j}{m}{m'}{j_1}{\!m_1}{\!m'_1}{j_2}{m_2}{m'_2}}
{(2j+1)(-1)^{j+m} (-1)^{j+m'} } = 
\frac{\hcoeff{j_2,}{-m_2,}{-m'_2}{j_1,}{\!m_1,}{\!m'_1}{j,}{-m,}{-m'}}
{(2j_2+1)(-1)^{j_2-m_2} (-1)^{j_2-m'_2}}
\end{equation}

Similarly, the 4D hyperspherical harmonics satisfy the following symmetry sign-inversion relation~\cite{Varshalovich1987}

\begin{equation}
U^{j}_{m,m'}  = (-1)^{m'-m} (U^{j}_{-m,-m'})^*
\end{equation}

It follows from linearity of the inner product that the same relation holds for the expansion coefficients $u^{j}_{m,m'}$.  Substituting both of these symmetry relations in Eq.~\ref{eqn:bispectrum} gives,

\begin{equation}
B_{j_1,j_2,j}  = (-1)^\epsilon \frac{2 j + 1}{2 j_2 + 1}\\
\sum_{m,m'}^{j} u^{j}_{-m,-m'} \sum_{m_1,m'_1}^{j_1} \sum_{m_2,m'_2}^{j_2} 
\hcoeff{j_2,}{-m_2,}{-m'_2}{j_1,}{\!m_1,}{\!m'_1}{j,}{-m,}{-m'} 
u^{j_1}_{m_1,m'_1} (u^{j_2}_{-m_2,-m'_2})^*
\end{equation}

where the parity exponent $\epsilon$ is given by the following expression in integer and half-integer indices 

\begin{eqnarray}
\epsilon & = & (m'-m) + (m'_2-m_2) + (j+m) + (j+m') + (j_2-m_2) + ( j_2-m'_2) \nonumber \\
& = & 2(j+m') + 2(j_2- m_2)
\end{eqnarray}

Both of the quantities in parentheses are integers for all values of $j$, $m'$, $j_2$, and $m_2$ and so the factor $(-1)^\epsilon$ is unity.  Finally, reversing the order of the summations over $m_2$, $m'_2$, $m$, and $m'$ yields

\begin{equation}
B_{j_1,j_2,j}  = \frac{2 j + 1}{2 j_2 + 1}\\
\sum_{m_2,m'_2}^{j_2} (u^{j_2}_{m_2,m'_2})^*  \sum_{m_1,m'_1}^{j_1}\sum_{m,m'}^{j}  
\hcoeff{j_2}{m_2}{m'_2}{j_1}{\!m_1}{\!m'_1}{j}{m}{m'} 
u^{j_1}_{m_1,m'_1} u^{j}_{m,m'}
\end{equation}

Comparison with the original Eq.~\ref{eqn:bispectrum} we see that the nested sum on the right-hand side is $B_{j_1,j,j_2}$.  A similar result can be obtained by interchanging $j_1$ and $j$, and so we find that $B_{j_1,j_2,j}$, $B_{j,j_2,j_1}$, and $B_{j,j,j_2}$ are all related by

\begin{equation}
\label{eqn:symm2}
\frac{B_{j_1,j_2,j}}{2j+1}  = \frac{B_{j,j_2,j_1}}{2 j_1+1}  = \frac{B_{j_1,j,j_2}}{2 j_2+1}.
\end{equation}

In addition to eliminating redundant bispectrum components, this relation greatly simplifies the calculation of forces.  In the original calculation, the spatial derivative of the bispectrum components was written as

\begin{eqnarray}
\label{eqn:dbisold}
\nabla B_{j_1,j_2,j}  & = & 
\sum_{m,m'}^{j} (\nabla u^j_{m,m'})^*\sum_{m_1,m'_1}^{j_1} 
\hcoeff{j}{m}{m'}{j_1}{\!m_1}{\!m'_1}{j_2}{m_2}{m'_2}
u^{j_1}_{m_1,m'_1} u^{j_2}_{m_2,m'_2} \nonumber \\
& & + \sum_{m,m'}^{j} (u^j_{m,m'})^*\sum_{m_1,m'_1}^{j_1} 
\hcoeff{j}{m}{m'}{j_1}{\!m_1}{\!m'_1}{j_2}{m_2}{m'_2}
\nabla u^{j_1}_{m_1,m'_1} u^{j_2}_{m_2,m'_2} \nonumber \\
& & + \sum_{m,m'}^{j} (u^j_{m,m'})^*\sum_{m_1,m'_1}^{j_1} 
\hcoeff{j}{m}{m'}{j_1}{\!m_1}{\!m'_1}{j_2}{m_2}{m'_2}
u^{j_1}_{m_1,m'_1} \nabla u^{j_2}_{m_2,m'_2} 
\end{eqnarray}

where the symbol $\nabla$ denotes the derivative of what follows with respect to the position of some neighbor atom.  We have dropped the double summation over $m_2$ and $m'_2$, as the coupling coefficients are non-zero only for $m_2 = m-m_1$, likewise for $m'_2$.  In the first term, the inner double sum over $m_1$ and $m'_1$ contains no derivatives, and so can be pre-calculated.  Hence, for the highest order bispectrum component ($j_1=j_2=j=J$), the computational complexity of this term is $O(J^2)$.  However, in the second and third term, the inner double sums contain derivatives and so must be calculated separately for each neighbor atom and for each entry in the outer double sums over $m$ and $m'$.  As a result, the computation complexity  of the second and third terms is $O(J^4)$.   Using the
symmetry relation given by Eq.~\ref{eqn:symm}, we can rewrite this as 

\begin{eqnarray}
\label{eqn:dbisnew}
\nabla B_{j_1,j_2,j}  & = & \hspace{0.24in}
\sum_{m,m'}^{j} (\nabla u^j_{m,m'})^*\sum_{m_1,m'_1}^{j_1} 
\hcoeff{j}{m}{m'}{j_1}{\!m_1}{\!m'_1}{j_2}{m_2}{m'_2}
u^{j_1}_{m_1,m'_1} u^{j_2}_{m_2,m'_2} \\
& & \hspace{-0.5in} + \frac{2 j + 1}{2 j_1 + 1} \sum_{m_1,m'_1}^{j_1} (\nabla u^{j_1}_{m_1,m'_1})^*\sum_{m,m'}^{j} 
\hcoeff{j_1}{m_1}{m'_1}{j}{\!m}{\!m'}{j_2}{m_2}{m'_2}
u^{j}_{m,m'} u^{j_2}_{m_2,m'_2} \nonumber \\
& & \hspace{-0.5in} + \frac{2 j + 1}{2 j_2 + 1} \sum_{m_2,m'_2}^{j_2} (\nabla u^{j_2}_{m_2,m'_2})^*\sum_{m_1,m'_1}^{j_1} 
\hcoeff{j_2}{m_2}{m'_2}{j_1}{\!m_1}{\!m'_1}{j}{m}{m'} 
u^{j_1}_{m_1,m'_1} u^{j}_{m,m'} \nonumber
\end{eqnarray}
  
Written in this way, all of the inner double sums are free of derivatives and can be pre-calculated. This has the effect of reducing overall the computational complexity from $O(J^4)$ to $O(J^2)$. For typical cases this results in more than one order of magnitude reduction in computational cost.

\end{document}